\documentstyle[11pt,newpasp,twoside,epsfig]{article}
\def\vs{\vskip 5pt}

\def\eg{{e.g., }}
\def\ie{{i.e., }}
\def\etal{{et al., }}

\def\etc{{etc.}}

\def\'{^{\prime}}

\def\avrg#1{{\langle #1 \rangle}}

\def\hmpc{{\, {\rm h}^{-1}~\rm Mpc}}

\def\kpc{{\rm~kpc}}

\def\msun{{\,M_\odot}}

\def\spose#1{\hbox to 0pt{#1\hss}}
\def\lta{\mathrel{\spose{\lower 3pt\hbox{$\mathchar"218$}}
     \raise 2.0pt\hbox{$\mathchar"13C$}}}
\def\gta{\mathrel{\spose{\lower 3pt\hbox{$\mathchar"218$}}
     \raise 2.0pt\hbox{$\mathchar"13E$}}}
\def\ge{\mathrel{\spose{\lower 3pt\hbox{$-$}}
     \raise 2.0pt\hbox{$\mathchar"13E$}}}
\def\le{\mathrel{\spose{\lower 3pt\hbox{$-$}}
     \raise 2.0pt\hbox{$\mathchar"13C$}}}

\begin{document}

\title{Concluding Remarks: From Current CMBology to its Polarization and
Sunyaev-Zeldovich Frontiers {\it circa} TAW8}
\author{J. Richard Bond}
\affil{Canadian Institute for Theoretical Astrophysics, University of Toronto, Toronto, Ontario, M5S 3H8, Canada}
\begin{abstract}
I highlight the remarkable advances in the past few years in CMB
research on total primary anisotropies, in determining the power
spectrum, deriving cosmological parameters from it, and more generally
lending credence to the basic inflation-based paradigm for cosmic
structure formation, with a flat geometry, substantial dark matter and
dark energy, baryonic density in good accord with that from
nucleosynthesis, and a nearly scale invariant initial fluctuation
spectrum. Some parameters are nearly degenerate with others and CMB
polarization and many non-CMB probes are needed to determine them,
even within the paradigm. Such probes and their tools were the theme
of the TAW8 meeting: our grand future of CMB polarization, with AMiBA,
ACBAR, B2K2, CBI, COMPASS, CUPMAP, DASI, MAP, MAXIPOL, PIQUE, Planck,
POLAR, Polatron, QUEST, Sport/BaRSport, and of Sunyaev-Zeldovich
experiments, also using an array of platforms and detectors, \eg
AMiBA, AMI (Ryle+), CBI, CARMA (OVROmm+BIMA), MINT, SZA, BOLOCAM+CSO,
LMT, ACT.  The SZ probe will be informed and augmented by new
ambitious attacks on other cluster-system observables discussed at
TAW8: X-ray, optical, weak lensing. Interpreting the mix is
complicated by such issues as entropy injection, inhomogeneity,
non-sphericity, non-equilibrium, and these effects must be sorted out
for the cluster system to contribute to ``high precision cosmology'',
especially the quintessential physics of the dark energy that adds
further mystery to a dark matter dominated Universe. We will have to
address ``Is it cluster evolution or is it cosmology?''. The answer
will be both, but we can be optimistic that, with the huge data
influx, computational power increase, and talented people joining the
adventure, we can handle both observationally, theoretically and
phenomenologically.
\end{abstract}

\section{Concordance? and its Consequences}

\subsection{The Beginning of the End or the End of the Beginning?}

\begin{figure}
\plotone{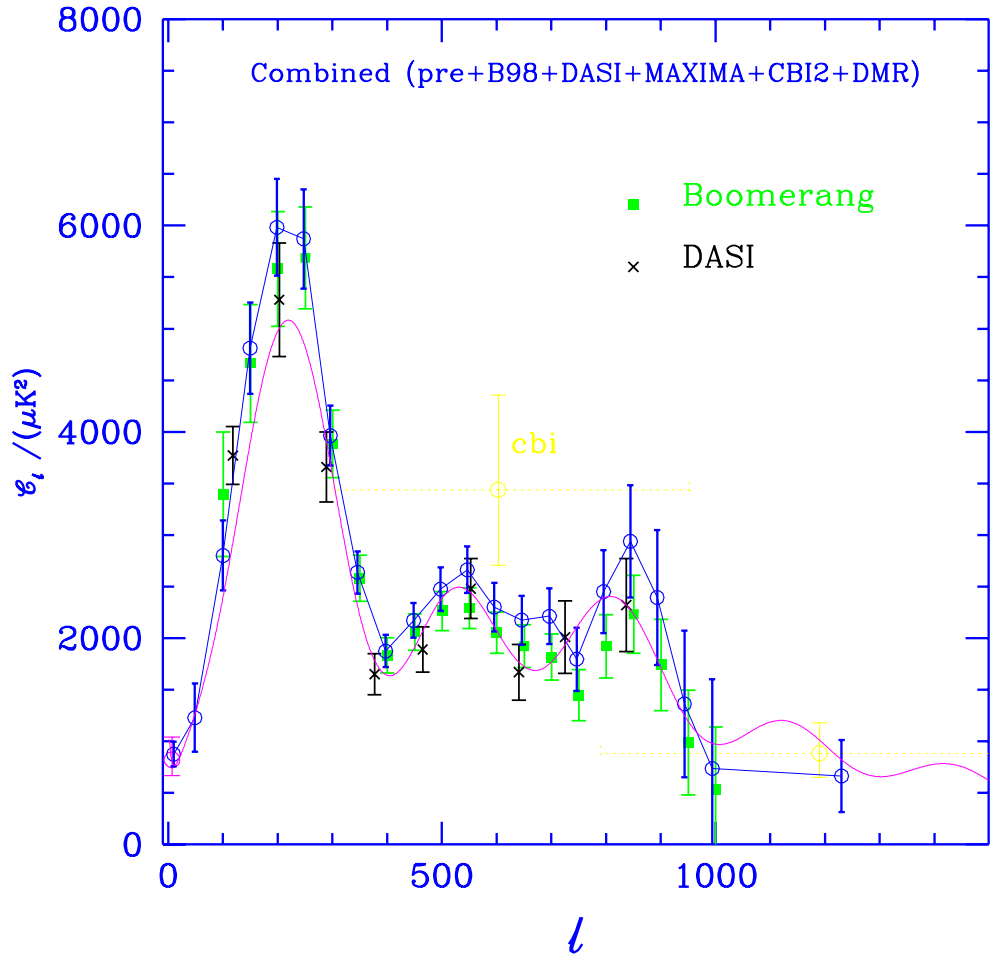}
\vspace{-5pt} \caption{The optimally-combined power spectrum ${\cal
C}_\ell$ grouped in bandpowers using all current data (circles, joined
by a light line) is contrasted with that for Boomerang-LDB (squares),
DASI (crosses) and DMR (point at low $\ell$). ``pre'' denotes TOCO,
Boom-97 and 19 other experiments predating April'99. This
heterogeneous ``prior CMB'' mix is quite consistent with what
Boomerang, DASI and Maxima show, with much larger errors. CBI2 denotes
the two published CBI points, only a small fraction of the total CBI
data.  A Boomerang best-fit model using the weakH+LSS+flatU prior is
also shown. In spite of the 10\% calibration and 13\% beam
uncertainties for Boomerang, little adjustment of its median values
was required by the other data. A caveat: DASI's fields 
overlap about 5\% of the Boomerang area, so there is correlation
between Boomerang and DASI. This is not taken into account here, but
the consistency in the overlap regions are currently being
explored. The optimal ${\cal C}_\ell$ without DASI included looks 
similar to the one shown, as might be expected given the consistency
of the two power spectra (and also the similarity in derived cosmic
parameters).  }
\label{fig:CLopt}
\end{figure}

In April 2001, just predating TAW8\footnote{This paper blends an
introductory primary CMB talk with my conference summary. Only a few
CMB references are given, organized by date (April'99, April'00,
April'01). The perpetrators of the advances mentioned in this summary,
and associated references, can be found elsewhere in these
proceedings.}, the Boomerang and DASI teams independently unveiled
remarkably similar power spectra of the {\it primary} anisotropies of
the CMB, those which can be calculated using linear perturbation
theory (Fig.~\ref{fig:CLopt}). The analysis of cosmological parameters
was in accord with indications from Large Scale Structure (LSS),
Supernova (SN1), and a variety of other observations, pointing towards
everyone's neo-standard model at this meeting, $\Lambda$CDM. Typical
$\Lambda$CDM parameters are taken to be: $\Omega_{tot}=1$;
$\Omega_\Lambda \approx 0.7$; Hubble parameter $h \approx 0.7$ from
the Hubble key project; $\Omega_m \sim 0.3$, including $\sim 0.04$ in
baryons, the rest in cold dark matter; $n_s$=1 as the slope of the
initial density power spectrum, the scale-invariant
Harrison-Zeldovich-Peebles value; overall mass density power
normalized to have $\sigma_8 \sim 0.9$, with $\sigma_8$ the {\it rms}
(linear) density fluctuation level on a cluster-scale ($8\hmpc$). The
baryon density choice $\Omega_{b}h^2 \approx 0.02$ is the Big Bang
Nucleosynthesis result calibrated with the deuterium abundance
estimated from absorption lines in QSO spectra.

\begin{figure}
\plotone{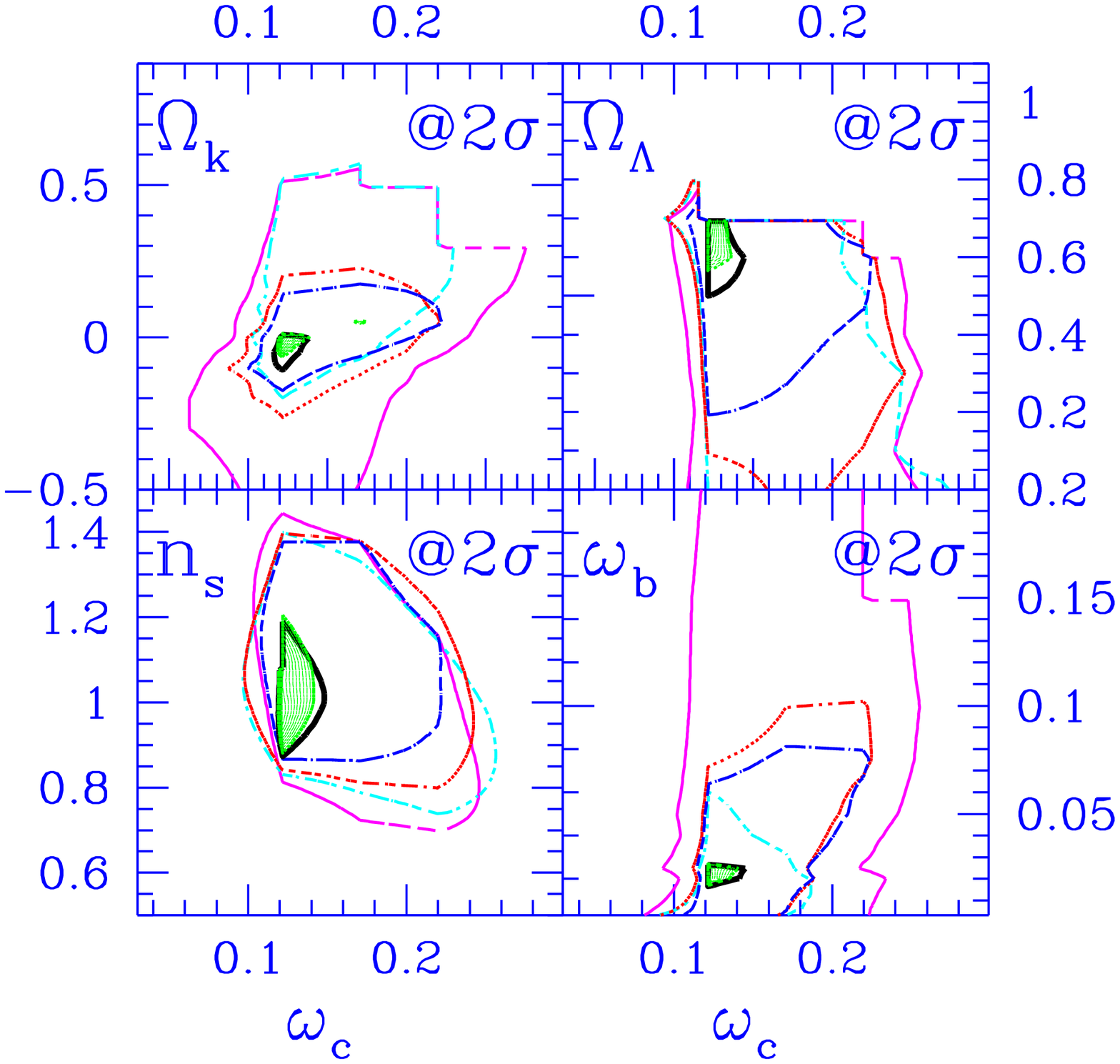}
\vspace{-10pt} \caption{2-$\sigma$ likelihood contours for the dark
matter density $\omega_c =\Omega_{cdm}{\rm h}^2$ and
$\{\Omega_k,\Omega_\Lambda,n_s,\omega_b\}$ for the LSS+weakH prior,
and the following CMB experimental combinations: DMR (short-dash); the
``April'99"+DMR data (short-dash long-dash); TOCO + (April'99+DMR)
data (dot short-dash); ``prior-CMB" = Boom-97 + (TOCO+April'99+DMR)
data (dot long-dash); Boomerang + DASI + Maxima-1 + ``prior-CMB" data 
(heavy solid, all-CMB). These $2\sigma$ lines tend to go from outside
to inside as more CMB experiments are added. The smallest 2-$\sigma$
region (dotted and interior) shows SN1+LSS+weakH+all-CMB, when SNI
data is added. For the $\Omega_\Lambda$, $n_s$ and $\omega_b$ plots,
the flatU prior, $\Omega_{tot}$=1, has also been assumed, but the
values do not change that much if $\Omega_{tot}$ floats. The main
movement from Apr'00 to Apr'01 was that $\omega_c$ localized more
around 0.13 in all panels, and the $\omega_b$ contour in the lower
right panel migrated downward a bit to be in its current good
agreement with Big Bang Nucleosynthesis. }
\label{fig:cmbLSS2sig}
\end{figure}

From the early 80s onward, CMB observations were used with LSS
information, as embodied in angular and redshift galaxy surveys,
cluster and other rare event abundances, cluster clustering, and
velocity flows to constrain the cosmological parameters defining the
space. Even in the days of CMB upper limits predating the
COBE/FIRS/SP91/Tenerife and subsequent detections, the CMB was a
powerful constrainer. When the COBE detection was combined with LSS, a
great collapse occurred in parameter space, which was further
constricted by detections on intermediate angular scales throughout
the 90s, and which Boomerang, DASI and Maxima have now turned into
bulls-eye determinations on some key parameters
(Fig.~\ref{fig:cmbLSS2sig}), focussing even more than in the April'00
release.

It appears from Fig.~\ref{fig:CLopt} that multiple peaks and dips in
the CMB have been found -- a dominant first peak, a less prominent
second one, and a hint of a third one, with interleaving dips
(April'01). These are even in roughly the right location of a
long-standing prediction of adiabatic inflation-based models with
little mean curvature. The physics of the ${\cal C}_\ell$ peak
structure is based on acoustic oscillations and velocity flows as the
photon-baryon fluid viscously passed from tight coupling to
free-streaming at photon decoupling (redshift $z_{dec} \sim 1100$,
about 0.4 Myr after the ``Big Bang''), generating the ``damping tail''
evident in the ``best-fit'' theoretical model shown in
Fig.~\ref{fig:CLopt}.  

The maps from which the ${\cal C}_\ell$ bandpowers are derived are
largely noise-free images of soundwave patterns seen through the
photon decoupling "surface" of width $\sim 10 \hmpc$ that defines the
thick-to-thin transition. This isquite a bit smaller than the comoving
"sound crossing distance" at decoupling, $\sim 100 \hmpc$ (\ie $\sim
100 \kpc$ physical), below which density oscillations and velocity
flows can be observed. After, photons freely-streamed along geodesics
to us, mapping (through the angular diameter distance relation) the
post-decoupling spatial structures in the temperature to the angular
patterns we observe now. The free-streaming along our (linearly perturbed)
past light cone leaves the pattern largely unaffected, except that
temporal evolution in the gravitational potential wells as the photons
propagate through them leaves a further $\Delta T$ imprint, called the
integrated Sachs-Wolfe effect.

Of course there are a number of other signals that are also present in
the maps, so how can we be confident that Fig.~\ref{fig:CLopt} really
offers a glimpse of fluctuation power at $z_{dec}$?  Known
contaminating signals include the Galactic foregrounds of
bremsstrahlung, synchrotron and dust emission, extragalactic radio and
infrared sources. As well, {\it secondary} anisotropies associated
with post-decoupling nonlinear effects are also present, These include
weak-lensing by intervening mass, Thompson-scattering by the nonlinear
flowing gas once it became "reionized" at $z \sim 10-20$, the thermal
and kinematic SZ effects, and the red-shifted emission from dusty
galaxies. All secondary and foreground sources leave non-Gaussian
imprints on the CMB sky, and all but the kinematic SZ effect have
different spectral signatures to aid in signal separation. For some
experiments (DASI, CBI), it has been crucial to remove sources, for
others like Boomerang, relatively contamination-free channels and
regions can be found.  We have been lucky that many of these
signals are subdominant at the angular scales we are probing in
Fig.~\ref{fig:CLopt}. As precision improves, signal separation
will loom large.

Because of the CMB+LSS success, we did not see at TAW8 as many of the
usual comparison cosmologies as we used to at such meetings, the open
oCDM, the hot/cold hybrid HCDM, $\tau$CDM, the tilted tCDM, the
cluster-normalized old-standard sCDM. Nor were cosmic defect models in
evidence. Though many of the $x$CDM's may have fallen away, we now see
$Q$CDM appearing on the stage, with $Q$ an ultra-low mass scalar
field, often called quintessence, that dominates at late times. Thus
$\Omega_Q$ replaces $\Omega_\Lambda$, and an effective $Q$-dynamics is
cast (though none too well) in terms of a mean pressure-to-density
ratio $w_Q = \bar{p}_Q/\bar{\rho}_Q$, an effective equation of state
(EOS). Well not so effective, since $Q$ is a spatially-dependent
field, or may be. In spite of a huge number of quintessential papers,
$Q$ would better stand for question mark. For $\Lambda$, $w_Q=-1$, but
$w_Q < - 1/3$ would get our patch of the Universe into acceleration,
apparently with no new comoving space to be revealed.

If there really is a $\Lambda$CDM/$Q$CDM concordance, then apart from
the wide grins of the ``often in error, never in doubt'' cosmologists,
hubrous abounding, we may also hear theorists' lament: Where are the
anomalies for wild and fun theorizing? Between our state now, with its
large-ish error bars and the never-ending worry about the systematic
rather than the statistical, and with the exquisite data from a vast
array of experiments coming down the pipe, there is still much room
for a cosmic surprise. Perhaps the greatest of all will be if the
models of the 80's do in fact describe how all of the structure formed
in the Universe, albeit with a mysterious dark energy accelerating
us. Even if $\Lambda$CDM, theorists are still at play, though not so
much at TAW8 which was concretely directed to the empirical. Just look
to the dark energy, the struggles to tie the latest inflation our
observable patch of the Universe now seems to be caught in with the
early inflation needed to ``smooth the universe'' and solve causality
problems, and, incidentally, to generate quantum noise from which all
observed cosmic structure originated. Look to the dialogues between
those of the M-theory brane worlds and the physical cosmologists,
reigniting the early universe connection that we were in danger of
losing -- what with the (clustering) dark matter being supposed cold
for so long and with inflation being generic but tunable to meet all
demands (though not without highly baroque additions).

\subsection{Broad Truths from the CMB+LSS}

Most amazing about Fig.~\ref{fig:CLopt}, COBE's FIRAS experiment, the
accumulating LSS information, now coming in a torrent with 2dF, Sloan
and higher redshift surveys, \etc, is that the paradigm appears to
hold: a hot Big Bang, with an almost perfect $T_{\gamma *} =2.725 \pm
0.001$K blackbody spectrum that must have come to us from beyond the
most distance SZ cluster, $z \gta 1$. That ${\cal C}_\ell$ is
significantly positive at $\ell \sim 1000$ argues against a large
$\exp[-2\tau_C]$ damping multiplier, where $\tau_C \sim 0.1
(\omega_b/0.02) (\omega_m/0.15)^{-1/2} ((1+z_{reh})/15)^{3/2}$ is the
Thompson optical depth to the epoch $z_{reh}$ of reheating. Thus,
though much pregalactic energy injection at $z \sim 200$ is still
possible, it does not look like it. The FIRAS limit of $4\bar{y} <
10^{-4}$ on fractional energy input into the CMB from the lack of a
Compton cooling spectral $y$-distortion further implies no large
entropy injection could have occurred at lower $z$ into the gas,
strongly limiting the role explosions can have had in LSS
development. The beautful direct connection of the small $\Delta T$
fluctuations to the density amplitudes now -- on the same spatial
scales -- strongly support the gravitational instability picture of
structure formation. That it forms hierarchically, from small to big,
is of course obvious from LSS observations at various redshifts, but
$n_s \sim 1 $ from the CMB adds further positive support.

The primary CMB fluctuations are quite Gaussian, according to COBE,
Maxima, and now Boomerang analyses. A non-Gaussian component, possibly
subdominant, of the primordial fluctuations can still work, but it is
encouraging for inflationists where Gaussian statistics are the
natural (but not only) outcome. Cosmic defect and cosmic string models
of structure formation are more challenged by the peaks and dips of
${\cal C}_\ell$, which are very difficult to get, than by the
Gaussianity.

We know the gravitational instability of a hierarchical Gaussian
random density field leads naturally to the cosmic web
interconnections and the prevalence of superclustering that we seem to
find observationally at low and high redshifts --- a framework for
thinking about the cluster/group system that was a main theme of
TAW8. The web consists of massive clusters with overdensities $\delta
\gta 100$, filaments with $\delta \sim 5-10$, which bridge massive
clusters, groups which bead the bridges, membranes with $\delta \sim
2$ which join the filaments, and the voids with $\delta < 0$
dominating the space but not the mass. This picture is of course borne
out by all the large simulations reported at TAW8, sizes ranging from
$128^3$ for a ``Schrodinger equation'' cosmological calculation to
$256^3$ and even $512^3$ for hydro and $N$-body, and to $1000^3$ for
$N$-body. Just a decade ago, a $128^3$ $N$-body was a tour de
force. We also heard much about semi-analytic methods in many
different guises that fit into this web picture, the halo model and
the peak-patch model with clustering included in both, and of course
many variants of Press-Schechter-ism.

\section{Using the Cluster/Group System and LSS to Probe a $\Lambda$ U}

Determining the dark energy EOS is the new mantra for the empirical
component of our subject, and because it turns out that CMB cannot
determine it by itself (unavoidable near-degeneracies exist among
cosmological parameters, $w_Q$ in particular), it will keep all cosmic
probers in business, probably for a very long time, all in the cause
of ``breaking degeneracies'', an oft-repeated phrase at TAW8. No
longer will the target be whether it is curvature energy $\Omega_k = 1
- \Omega_{tot}$ or $\Omega_\Lambda$ that makes up the deficit between
$\Omega_m$ and unity, rather it is the much subtler and harder EOS
(and more refined) dynamics that we must use our probes to
determine. High redshift supernovae, to be sure, will be used in large
surveys, but also: weak lensing of large scale structure and cluster
abundances as a function of redshift, informed by Sunyaev-Zeldovich,
optical, X-ray and lensing surveys, possibly group and galaxy
evolution, \ie the themes and ambitious plans expressed at TAW8. And
though we know the sad history of how the classic grand cosmological
tests of the deceleration parameter ran afoul of whether it is the
``messy astrophysics of complex evolving objects'' or ``cosmology'',
we have little choice but to understand our systems well enough so
they too can become parameter-estimation tools.  How else but through
astronomy can we learn about perhaps the greatest mystery in all of
physics?

\vs
\noindent 
{\bf Clusters are Not Simple:} When the differences we were going for
were vast (cluster abundances as a function of redshift for $\tau$CDM
or sCDM {\it cf.}  $\Lambda$CDM, given normalization to clusters now),
one could be slightly cavalier about the complexity of clusters -- the
deepest potential wells in the known universe, nice equilibrium
systems.  That is, theoretical naivete could be forgiven. Even the use
of the Press-Schechter mass functions, $\beta$-models, isothermality,
spherical profiles, single-phase assumptions, ignoring the known
complications of magnetic fields, cooling flows, metal/energy
injections, and the emerging complications revealed by the new Chandra
and XMM data, could be forgiven as long as great accuracy was not
claimed. 

I believe success in the dark energy EOS enterprise using clusters is
possible, for many reasons in evidence at TAW8: the
X/optical/lensing/SZ cluster information here now and planned; the
overwhelming avalanche of high quality survey data to come,
terapixels-worth; the ambitious theoretical work using hydrodynamics,
N-body, analytic and semi-analytic tools being undertaken to
understand the data and also forecast and prepare for future
experiments; the computing horsepower that promises Monte Carlo
simulation to take theory fully forward into the observational space
-- with theorists becoming fully integrating into the
experimental/observational teams; and especially with the energetic
young researchers avidly embracing the complexity.

On the other hand, clusters in the X-ray at higher resolution do not
look simple, red galaxy numbers per cluster mass must fluctuate,
merging at $z \sim 1$ will be ubiquitous, so equilibrium may not
prevail, especially in the most interesting objects that catch our
various ``eyes''.

\vs
\noindent 
{\bf Some Clusters May Not Be Too Complex:} Armed with all of the
probes and surveys at our disposal, we should be able to select
physically-understood cluster subsamples for which we can be
reasonably sure that the cosmic parameters we deliver will be with 
calculable systematic errors and no bias in value, not just with the
small statistical errors that naive theory forecasts. Of course this
is preaching to the converted. Given the range of talks at TAW8,
almost all terrain we need to cover was covered at some level:

\begin{itemize}

\item Metals in the intracluster/intragroup medium, and in the higher
$z$ intergalactic medium ($\lta 20\%$ apparently affected at $z \sim
3$), relating to, but far from solving, the major issue of energy
injection into these media. 

\item Filaments may be more SZ-observable if the energy injection is
strong. Related questions of the effects feedback has on group and
cluster gas probes as a function of redshift remain unanswered.

\item Great plans for SZ-interferometer surveys: AMiBA is developing
MMIC HEMTs at 90 Ghz, novel correlators, platform, \etc\, with survey
plans to probe deep, medium and shallow, with coverage 3, 70 and 175
sq deg. The CARMA integration of the BIMA and the OVRO mm arrays, both
of which have had a spectacular history already in SZ science, the
CBI, targeting $z< 0.1$ clusters, the SZA at 30 GHz, AMI at 15
GHz and MINT at 140 GHz will all considerably enhance the SZ effort. 

\item Great plans for bolometer-based surveys: the CSO with BOLOCAM on
Mauna Kea soon to observe, ACBAR at the South Pole already observing,
SuZIE of course, the LMT (large mm telescope) in Mexico, eventually
Planck. There is much excitement about bolometer arrays on
ground-based $\sim 6m$ telescopes, \eg the ACT proposal for 3 32x32
pixel bolometer arrays delivering $1.7^\prime$ resolution. 

\item Progress in analysis pipelines for of all of the different types
of data that is coming, though much remains to do. For SZ, component
signal separation and source identifications are crucial. It is ironic
that the primary CMB, so long our target, is a nuisance confusion to
be filtered out. For optical spectra, Principal Component Analysis was
effectively used.  Only 3 eigenmodes describing old, field and
post-Star-Formation spectra were needed, and helped clarify cluster
gradients and the Butcher/Oemler effect. 

\item The intense work in the optical on clusters and groups, both for
specific objects and in heroic surveys. The now venerable CNOC1,2. The
100 sq deg ``Red Sequence Cluster Survey'', with its 22 patches of 5
sq deg, can deliver optically-identified clusters in abundance: 200 at
$z>1$, 500 at $ z>0.7$, 2700 at $z>0.4$. More ambitious areas are
planned: RCS2's 1000 sq deg; VISTA's 10000 sq deg; Megacam on CFHT 9
sq deg/night, applied to the CFH Legacy Survey. By contrast, SDSS
though wide is relatively shallow, with the clusters dying off above
$z \sim 0.5$. The 130 sq deg Las Campanas survey used a 1 m telescope to
get clusters in the $0.35 <z< 0.9$, with extensive follow-up, with
application to the key LSS cluster clustering $r_0-d_c$ figure.

\item Groups of $\sim 10^{13-14} \msun$ and poor clusters are such a
mix, making detection ambiguous, prone to superposition. Still, they
are the {\it rms} objects in the universe, so we must understand them,
and there is some progress there on selected populations.

\item The new substantive X-ray luminosity functions at different
redshifts, BCS, ROSSI, REPLEX, EMSS, SHARC, NEP, MACS -- do I have them
all? -- giving a consistent picture it seems. Would that the relation
of $L_X$ to mass was simple, $n(T_X)$ much preferred, of course, but
the clusters that can be used are still small in number. Still, 
the $\Lambda$CDM concordance model seems to work yet again.

\item The X-ray studies of individual clusters revealing finer detail
(subarcserc resolution for Chandra) and complexity in the intracluster
medium (temperature inhomogeneity, ``cold fronts'' {et al.}, and, as
in all such meetings, cooling flows). Can we cosmic hydro simulators
handle this richness of detail?

\item The beautiful lensing mass maps of individual clusters, and a
supercluster example, and the amazing strongly-lensed clusters with
powerful and multiple arcs at high $z$; and galaxy-galaxy lensing too.

\item The weak-lensing probe of LSS a tool to get beyond galaxy
biasing to the mass density power spectrum, and through that and
higher order non-Gaussian statistics, to cosmic parameters, including
$w_Q$.  

\item The heavy use of new instruments (Subaru figured prominently
here, as of course did Chandra and XMM), the use of venerable
telescopes, sometimes newly instrumented (optical surveys at CFHT,
CTIO, {\it etc.}), and an imaginative panoramic optical imager, an
array of small telescopes for weak lensing and other mappings.

\end{itemize}

\section{The Primary CMB Snapshot and the Race to Polarization}

\subsection{The Recent Primary CMB Experiments} \label{sec:CMBsnapshot}

Fig.~\ref{fig:CLopt} gives the current snapshot of our knowledge of
the power spectrum ${\cal C}_\ell$ $\equiv \ell(\ell+1)\avrg{\vert
(\Delta T)_{\ell m}\vert^2}/(2\pi)$ as a function of multipole $\ell$
in a spherical harmonic expansion $(\Delta T)_{\ell m}$ of {\it
primary} total temperature anisotropies. All published CMB
experimental results as of Fall 2001, including their quoted
calibration and beam errors, are compressed into 22 bandpowers:

\begin{itemize}

\item BOOMERanG-98, a long duration balloon (LDB) experiment took a
1.2m telescope aloft from McMurdo Bay in Antarctica in late Dec~1998
with 16 bolometers cooled to 300 mK at frequencies 90, 150, 220 and
the dust-dominated 400 GHz. It circled the Pole for 10.6 days, mapping
1800 sq degs with a best resolution of $10.7^\prime$ (Gaussian $\ell_s
\approx 750$). For April'01 (the results shown here), 800 sq deg and
four of five 150 GHz bolometers were used, about eight times more data
than was used for April'00.  We are now analyzing $\sim 1300$ sq degs
with $3.5^\prime$ pixels. A 10\% calibration and a 13\% beam
uncertainty must be included.

\item MAXIMA-I, a short duration (overnight) balloon, used bolometers
cooled to 100 mK to map 124 sq deg to $\sim 10^\prime$. There was a
4\% calibration and a 5\% beam uncertainty.

\item DASI, the South-Pole-based 13 element (0.2m antennae) Degree Angular
Scale Interferometer with 30 GHz HEMTs,  mapped 288 sq deg in 32 fields of 3.4
deg diameter over the $\ell$-range 100-900. There was a 4\%
calibration uncertainty, but none in the beam.

\item CBI, the Chile-based 13 element (0.9m dish) Cosmic Background
Imager interferometer with 30 GHz HEMTs, has mapped three 10 sq deg
mosaic regions and three 0.44 sq deg deep fields in 2000-01, probing
from $\ell \sim 300$ up to $\sim 4000$, \ie well beyond the Boomerang
range into the ``damping tail''. There was a 3\% calibration
uncertainty, but again none in the beam. CBI2 denotes the Nov'00 CBI
bandpowers that used two of the deep fields and only 5\% of the total
data in the analysis. The CBI team is collaborating with our group at
CITA in a much more extensive analysis of the year-2000 CBI mosaic and
deep field data, which will significantly sharpen the focus in the
$\ell \gta 1000$ regime ($\sim$ Jan'02 release).

\item Boom-97, the North American test flight.
 
\item TOCO, a Chile-based telescope which used SIS as well as HEMT
receivers.

\item COBE-DMR, with resolution $\ell_s \approx 17$.  

\item April'99: 19 other earlier CMB experiments'' that had bandpowers
(or upper limits) we were using by April 99. ``prior CMB'' or ``pre''
adds TOCO and Boom-97 to the mix, \ie all CMB data before the April'00
Boomerang release.

\end{itemize}

The band positions and $\Delta \ell $=50 widths were chosen to be
those of the April'01 Boomerang release (Netterfield \etal 2001),
except a narrower first bin ($3\le \ell \le 25$) was added to
encompass the COBE DMR results and the $\ell > 1025$ region beyond the
Boomerang range, but encompassing CBI2, is much wider ($\Delta
\ell$=500).

\subsection{More on the Cast of Cosmic Parameters}

It has long been recognized that the measurement of the predicted
${\cal C}_\ell$ structures such as peaks and dips and damping tails
could determine cosmic parameters. The ``minimal'' set
$\{\Omega_{tot},\Omega_{b}h^2,\Omega_{cdm}h^2, \Omega_{hdm}h^2,
\Omega_{wdm}h^2, n_s , \sigma_8, \tau_C \}$ defined an operative
parameter space including hot, warm or cold dark matter, as well as
baryonic, from 1982 onwards. (We now prefer to use $\omega_j \equiv
\Omega_{j}h^2$ because it is related to the physical density rather
than a ratio to the critical density.) A target (scalar) spectral
index $n_s$ was the Harrison-Zeldovich-Peebles 1, but even in 1982,
nearly scale invariant emerged, with the tilt $n_s-1$ near to
unity. Cosmologists treated it as a free parameter.  We parameterized
the power amplitude in initial mass density fluctuations by a ``galaxy
biasing factor'' that was almost exactly $\sigma_8^{-1}$; $\sigma_8$
became the more widely adopted normalizer in 1985. Whether there was
early reheating, embodied in $\tau_C$, has always been a question.

The ancient $\Omega_{\Lambda}$, never as abhorent to cosmologists as
it was to particle physicists, was resurrected in the mid-80s under
the ``what you see is what you get'' mantra that Jim Peebles chanted
for us based on the annoying ubiquity of $\Omega_m <1 $. For me it
came into sharp focus in 1986, since $\Omega_{\Lambda}$ was one of the
ways within the inflation paradigm to help explain the large scale
power first seen in cluster clustering, then in velocity flows, then
in galaxy clustering. In response, we were actively considering
$x$CDM models that were: open (oCDM); hot/cold hybrids; high in
$\Omega_b/\Omega_m$ (BCDM); radically ``broken scale invariance''
cases, with hills and/or valleys in $n_s (k)$; tilted ($n_s \sim
0.6$); high in the density of relativistic decay products of
decaying keV-level neutrinos ($\tau$CDM); isocurvature, from quantum
noise in scalar fields other than the inflaton; adiabatic/isocurvature
hybrids. Even the venerable isocurvature baryon-dominated model of the
70s, with $\Omega_{tot}<1$ and $n_s$ far from scale invariance, was
resurrected, again by Peebles. 

In the mid-80s, it was also recognized that tensor modes in the
temperature fluctuations driven by gravitational wave zero-point
fluctuations are a natural consequence of inflation, expanding the
parameter space to include a relative tensor-to-scalar power
$\tilde{r}_{ts}$, and a tilt $n_t$ independent of $n_s$.

In the late-90s, in response to the $\Lambda$ mystery, the EOS
parameter $w_Q$ was added, and sometimes so was $\dot{w}_Q$, a measure
of its time variation.

\subsection{Zeroing in on the Cosmological Parameters} 

Fig.~\ref{fig:cmbLSS2sig} shows what happens to $\Omega_{tot}$,
$\Omega_\Lambda$, $\Omega_bh^2$, $\Omega_{cdm}h^2$ and $n_s$ in the
parameter space described below as results from the Sec.~\ref{sec:CMBsnapshot} CMB
experiments are added to LSS information on $\sigma_8$ (as estimated
from cluster abundances) and a density power spectrum shape/tilt
parameter $[\Gamma +(n_s-1)/2]$, with $\Gamma \propto \Omega_m h$ (as
estimated from large galaxy clustering surveys). The distributions in
both these LSS parameters were taken to be quite broad, reflecting our
desire to be uncontroversial among LSS practitioners. To this ``LSS
prior'' probability, a ``weakH prior'' was imposed, requiring that the
Hubble parameter and age of the universe satisfy $0.45 < h < 0.9$ and
$t_0>10$ Gyr. Given the emerging CMB localization of $\Omega_{tot}$
near unity, and the ``inflationist's theoretical prior'' of penalizing
the baroqueness which large mean curvature models with non-negligible
$\vert 1- \Omega_{tot}\vert $ suffer from, a ``flatU prior'' is
adopted in 3 of the 4 panels (although it only makes a notable
difference in the $\Omega_\Lambda$ panel). The innermost (filled)
$2\sigma$ contour of Fig.~\ref{fig:cmbLSS2sig} adds a ``SN1a prior''
to weakH+LSS, using the likelihood function in
$\Omega_\Lambda$-$\Omega_m$ space derived from the high redshift
Supernova 1a data.

With just the COBE-DMR+LSS data, the 2-$\sigma$ contours already
localize in $\Omega_{cdm}h^2$ thanks to LSS. Fig.~\ref{fig:cmbLSS2sig}
shows there is also some localization of $n_s$ around unity, and this
is true even without the LSS prior. Although the April'99 data
collectively shows evidence for a peak, it is not well enough
localized for a useful curvature constraint. The picture begins to
improve later in 1999, with $\Omega_k$ localizing near zero when TOCO
is added to the April'99 data. In April'00, results from the first CMB
LDB flight, Boomerang, were announced, followed by those from Maxima,
then the first CBI results in Nov'00. In April'01, Boomerang and DASI
announced compatible power spectra, and Maxima improved its power
spectrum with finer pixelization. Fig.~\ref{fig:cmbLSS2sig} looks
nearly identical if only Boomerang+DMR are used for the CMB
experiments, and DASI+DMR also looks very similar, except
$\Omega_{cdm}h^2$ is not quite as localized. $\omega_b$, $\omega_c$
and $\Omega_\Lambda$ really focus in with Boomerang.

 The data therefore favour the simplest (least baroque) inflation
theories: nearly flat, nearly scale invariant primordial fluctuations
nothing on gravitational waves yet. The baryon density is nearly the
Big Bang Nucleosynthesis value. CMB+LSS implies there is substantial
dark matter and dark energy. As well, there are derived quantities we
can get: Hubble constant ($56 \pm 9$) and age ($15 \pm 2$ Gyr) are
consistent ($62 \pm 6$, $14 \pm 1$ if $\Omega_{tot}=1$). It is
interesting to note that LSS data and CMB data are independently
pointing to some of the same values. For example, the 2dF survey finds
$\Omega_m h=0.20\pm .03$, $\Omega_b/\Omega_m=0.15 \pm .07$ {\it cf.}
our $0.21\pm .05, 0.14 \pm .03$. If a light massive neutrino is added
(H$\Lambda$CDM models), CMB does not discriminate, and CMB+LSS just
shifts things to slightly lower but still nonzero
$\Omega_\Lambda$. However, to actually find evidence for or against at
this stage, one needs LSS+CMB+SN1 to discriminate.

For the Quintessence EOS, we have found $w_Q < -0.3$ at $2\sigma$ for
the CMB with the LSS+flatU+weakH priors. It is only when the SN1 prior
is included that the $w_Q$ constraint, $<-0.7$, becomes rather
restrictive.

We are only at the beginning of the high precision CMB era. The
bolometer-based ACBAR and the Arkeops and Tophat LDBs already have
data, as does the HEMT-based interferometer VSA (Very Small Array) in
Tenerife. CBI and DASI continue to accumulate data. And NASA launched
the all-sky HEMT-based MAP satellite on June 30, 2001, with
$12^\prime$ resolution. It is now mapping the sky at L2, the second
Lagrangian point of the earth moon system some 1.5 million km
away. Further downstream, in 2007, ESA will launch the
bolometer+HEMT-based Planck satellite, with $5^\prime$ resolution.

How many independent cosmic parameter combinations can be measured
with the CMB now? Four linear combos were forecasted to be determined
to $\pm 0.1$ for Boomerang, and this is the result obtained in the
analysis of the real data. Adding the LSS-prior brought a 5th into
this precision level.  Our future involves the precision that all-sky
mapping can give: for prior-less CMB-only, 6/9 to $\pm$ .1 , 3/9 to
$\pm$ .01 for MAP, 7/9 to $\pm$ .1, 5/9 to $\pm$ .01 for
Planck. ($\Omega_{hdm}h^2$ and $\tilde{r}_{ts}$ are now added to our
basic mix of 7).

\subsection{Running or Planned Polarization Experiments}

Given the total ${\cal C}_\ell$ of Fig.~\ref{fig:CLopt}, we can
forecast what the polarization signal and its cross-correlation with
the total anisotropy will be, and which $\ell$ range gives the maximum
signal: $\sim 5 \mu$K over $\ell \sim 400-1600$ is a target for the
$E$-mode that scalar fluctuations give. We cannot yet forecast the
strength of the $B$-mode signal induced by gravity waves, since there
is as yet no evidence for or against them in the data. However, the
amplitude would be very small indeed even at $\ell \sim 100$ where it
is biggest: for now, detection is what theorists' dream of, because it
would tell us so much about inflation, but we have great faith in
the ingenuity of the experimentalists.

 A great race is on to first detect the $E$-mode: bets on which team?
The experiments discussed at TAW8, and listed in the abstract, range
from many degrees to subarcminute scales.  The PIQUE 95\% CL upper
limit of 14 $\mu$K at $\ell\sim 210$ and the similar ATCA limit (but
at $\ell\sim 4500$) are still well above the forecast, but we are
getting there. 

The target $\ell$ range and amplitude of the forecasts has, of course,
not escaped the notice of experimental designers;, \eg AMiBA, CBI with
HEMTs, ACBAR, Boomerang-2K2, and eventually QUEST and Planck, of
course, with polarization-sensitive bolometers, can all probe this
range. Not surprisingly, the forecasts show a solid detection is
likely for many of the proposed experiments, often with enough
well-determined broad-band powers to use the results for cosmology.
Since polarization is such a necessary outcome of the adiabatic
paradigm, the implications will be enormous (and exciting) if it fails
to be there.

What about the competition from foreground and secondary polarization?
For the SZ secondary, the contribution is small, but for Galactic
foregrounds it could be large -- not enough is known about them at the
CMB observing frequencies. If we can unravel the signals that make it
up the detections, the primary polarization will come into its own to
augment total anisotropy in cosmic parameter estimation. It can break
parameter degeneracies, \eg those associated with allowing $n_s(k)$
structure, by using the shift in the polarization ${\cal C}_\ell$
peaks and dips relative to those for the total anisotropy.

The quest for the polarization signal is sufficiently exciting that
the adventure travel offered, to the Atacama desert where the CBI is
(and ALMA will be), to the South Pole for a winter-over, to Hawaii or
California, is a bonus.

\acknowledgments I would like to thank my fellow CITAzen CMBologists 
and other Boomerang and CBI team members for an amazing trek over the
past few years as the ups and downs of ${\cal C}_\ell$ have been
progressively revealed.

{\it Thanks from all of us to Lin-Wen, Jackie, Jean \etal for a great
job of organizing us and to Fred and Pauchy for bringing us here. The
hospitality, the city/mountain split venue, the cuisine, and the
science, were all tremendous. Congratulations on the wonderful growth
of effort in Taiwan associated with CosPA. We look forward to AMiBA
science, and the many opportunites for the further meetings 
it will precipitate in our fast advancing subject.}

\end{document}